On the Bayesness, minimaxity and admissibility of point estimators of allelic frequencies

Carlos Alberto Martínez[1,2], Khitij Khare[2] and Mauricio A. Elzo[1]

[1] Department of Animal Sciences, University of Florida, Gainesville, FL, USA

[2] Department of Statistics, University of Florida, Gainesville, FL, USA

Correspondence: C. A. Martínez, Department of Animal Sciences, University of Florida, Gainesville, FL 32611, USA.

Tel: 352-328-1624.

Fax: 352-392-7851.

E-mail: carlosmn@ufl.edu

**Abstract**

In this paper, decision theory was used to derive Bayes and minimax decision rules to estimate allelic frequencies and to explore their admissibility. Decision rules with uniformly smallest risk usually do not exist and one approach to solve this problem is to use the Bayes principle and the minimax principle to find decision rules satisfying some general optimality criterion based on their risk functions. Two cases were considered, the simpler case of biallelic loci and the more complex case of multiallelic loci. For each locus, the sampling model was a multinomial distribution and the prior was a Beta (biallelic case) or a Dirichlet (multiallelic case) distribution. Three loss functions were considered: squared error loss (SEL), Kulback-Leibler loss (KLL) and quadratic error loss (QEL). Bayes estimators were derived under these three loss functions and were subsequently used to find minimax estimators using results from decision theory. The Bayes estimators obtained from SEL and KLL turned out to be the same. Under certain conditions, the Bayes estimator derived from QEL led to an admissible minimax estimator (which was also equal to the maximum likelihood estimator). The SEL also allowed finding admissible minimax estimators. Some estimators had uniformly smaller variance than the MLE and under suitable conditions the remaining estimators also satisfied this property. In addition to their statistical properties, the estimators derived here allow variation in allelic frequencies, which is closer to the reality of finite populations exposed to evolutionary forces.

## 1. Introduction

Allelic frequencies are used in several areas of quantitative and population genetics, hence the necessity of deriving point estimators with appealing statistical properties and biological soundness. They are typically estimated via maximum likelihood, and under this approach they are treated as unknown fixed parameters. However, Wright (1930; 1937) showed that under several scenarios allele frequencies had random variation and hence should be given a probability distribution. Under some of these scenarios, he found that the distribution of allele frequencies was Beta and that according to the particular situation its parameters had a genetic interpretation (Wright 1930; 1937; Kimura and Crow, 1970). For instance, under a recurrent mutation scenario, the parameters of the Beta distribution are functions of the effective population size and the mutation rates (Wright, 1937). Expressions for these hyperparameters under several biological scenarios and assumptions can be found in Wright, (1930; 1937) and Kimura and Crow (1970).

Under the decision theory framework, given a parameter space $\Theta$, a decision space $D$, observed data $\boldsymbol{X}$, and a loss function $L(\theta, \delta(\boldsymbol{X}))$, the average loss (hereinafter the frequentist risk or simply the risk) for a decision rule $\delta$ when the true state of nature is $\theta \in \Theta$, is defined as $R(\theta, \delta) = E_\theta[L(\theta, \delta(\boldsymbol{X}))]$. The ideal decision rule, is one having uniformly smallest risk, that is, it minimizes the risk for all $\theta \in \Theta$ (Lehmann and Casella, 1998). However, such a decision rule rarely exists unless restrictions like unbiasedness or invariance are posed over the estimators. Another approach is to allow all kind of estimators and to use an optimality criterion weaker than uniformly minimum risk. Such a criterion looks for minimization of $R(\theta, \delta)$ in some general sense and there are two principles to achieve that goal: the Bayes principle and the minimax principle (Lehman and Casella, 1998; Casella and Berger, 2002).

Given a loss function and a prior distribution, the Bayes principle looks for an estimator minimizing the Bayesian risk $r(\Lambda, \delta)$, that is, a decision rule $\delta^*$ is defined to be a Bayes decision rule with respect to a prior distribution $\Lambda$ if it satisfies

$$r(\Lambda, \delta^*) = \int_\Theta R(\theta, \delta^*) d\Lambda(\theta) = \inf_{\delta \in D} r(\Lambda, \delta).$$

This kind of estimators can be interpreted as those minimizing the posterior risk. On the other hand, the minimax principle consists of finding decision rules that minimize the supremum (over the parameter space) of the risk function (the worst scenario). Thus $\delta^*$ is said to be a minimax decision rule if

$$\sup_{\theta \in \Theta} R(\theta, \delta^*) = \inf_{\delta \in D} \sup_{\theta \in \Theta} R(\theta, \delta).$$

The aim of this study was to derive Bayes and minimax estimators of allele frequencies and to explore their admissibility under a decision theory framework.

## 2. Materials and methods

2.1 Derivation of Bayes rules

Hereinafter, Hardy-Weinberg equilibrium at every locus and linkage equilibrium among loci are assumed. Firstly, the case of a single biallelic locus is addressed. Let $X_1, X_2$ and $X_3$ be random variables indicating the number of individuals having genotypes AA, AB and BB following a trinomial distribution conditional on $\theta$ (the frequency of the "reference" allele B) with corresponding frequencies: $(1-\theta)^2, 2\theta(1-\theta)$ and $\theta^2$, and let $\boldsymbol{X} = (X_1, X_2, X_3)$. Therefore, the target is to estimate $\theta \in [0,1]$. Thus, in the following, the sampling model is a trinomial distribution and the prior is a Beta$(\alpha, \beta)$. This family of priors was chosen because of mathematical convenience, flexibility, and because as discussed previously, the hyperparameters $\alpha$ and $\beta$ have a genetic interpretation (Wright, 1937). Under this setting, three loss functions were used to derive Bayes decision rules: squared error loss (SEL), Kullback-Leibler loss (KLL) and quadratic error loss (QEL).

2.1.1 Squared error loss

Under SEL, the Bayes estimator is the posterior mean (Lehman and Casella, 1998; Casella and Berger, 2002). Thus we need to derive the posterior distribution of the parameter:

$$\pi(\theta|X) \propto \pi(X|\theta)\pi(\theta)$$

$$\propto (1-\theta)^{2x_1+x_2}\theta^{x_2+2x_3}\theta^{\alpha-1}(1-\theta)^{\beta-1}$$

$$= \theta^{x_2+2x_3+\alpha-1}(1-\theta)^{2x_1+x_2+\beta-1}.$$

Therefore, the posterior is a Beta($x_2 + 2x_3 + \alpha, 2x_1 + x_2 + \beta$) distribution and the Bayes estimator under the given prior and SEL is:

$$\hat{\theta}^{SEL} = \frac{x_2 + 2x_3 + \alpha}{x_2 + 2x_3 + \alpha + 2x_1 + x_2 + \beta}$$

$$= \frac{x_2 + 2x_3 + \alpha}{2n + \alpha + \beta} \quad (\because x_1 + x_2 + x_3 = n)$$

The frequentist risk of this estimator is:

$$R(\theta, \hat{\theta}^{SEL}) = E_\theta\left[\left(\frac{X_2 + 2X_3 + \alpha}{2n + \alpha + \beta} - \theta\right)^2\right]$$

$$= Var_\theta\left[\frac{X_2 + 2X_3 + \alpha}{2n + \alpha + \beta}\right] + \left(E_\theta\left[\frac{X_2 + 2X_3 + \alpha}{2n + \alpha + \beta} - \theta\right]\right)^2$$

$$= \frac{Var[X_2] + 4Var[X_3] + 4Cov[X_2, X_3]}{(2n + \alpha + \beta)^2} + \left(E_\theta\left[\frac{X_2 + 2X_3 + \alpha - \theta(2n + \alpha + \beta)}{2n + \alpha + \beta}\right]\right)^2.$$

Using the forms of means, variances, and covariances of the multinomial distribution yields:

$$R(\theta, \hat{\theta}^{SEL}) = \frac{2n\theta(1-\theta)(1-2\theta(1-\theta)) + 4n\theta^2(1-\theta^2) - 4n(2\theta(1-\theta)\theta^2)}{(2n+\alpha+\beta)^2}$$

$$+ \left(\frac{2n\theta(1-\theta) + 2n\theta^2 + \alpha - \theta(2n+\alpha+\beta)}{2n+\alpha+\beta}\right)^2$$

$$= \frac{2n\theta(1-\theta) + [\alpha(1-\theta) - \beta\theta]^2}{(2n+\alpha+\beta)^2}.$$

Note that the problem has been studied in terms of counts of individuals in each genotype, but it can be equivalently addressed in terms of counts of alleles. To see this, let $Y_1$ and $Y_2$ be random variables corresponding to the counts of B and A alleles in the population; consequently, $Y_1 = 2X_3 + X_2, Y_2 = 2X_1 + X_2$ and $Y_1 = 2n - Y_2$. Now let $\boldsymbol{Y} := (Y_1, Y_2)$; therefore, $\pi(\boldsymbol{Y}|\theta) \propto \theta^{y_1}(1-\theta)^{2n-y_1}$ a Binomial$(2n, \theta)$ distribution. With this sampling model and the same prior $\pi(\theta)$, $\pi(\theta|\boldsymbol{Y})$ is equivalent to $\pi(\theta|\boldsymbol{X})$ given the relationship between $\boldsymbol{Y}$ and $\boldsymbol{X}$. For the biallelic loci case, $\pi(\theta|\boldsymbol{X})$ will continue to be used. Notwithstanding, as will be discussed later, for the multi-allelic case working in terms of allele counts is simpler.

2.1.2 Kullback-Leibler loss

Under this loss, the Bayes decision rule is the one minimizing (with respect to $\delta$):

$$\int_0^1 L_{KL}(\theta, \delta)\pi(\theta|\boldsymbol{X})d\theta$$

where:

$$L_{KL}(\theta, \delta) = E_\theta\left[\ln\left(\frac{\pi(\boldsymbol{X}|\theta)}{\pi(\boldsymbol{X}|\delta)}\right)\right] = E_\theta\left[\ln\left(\frac{(1-\theta)^{2X_1+X_2}\theta^{X_2+2X_3}}{(1-\delta)^{2X_1+X_2}\delta^{X_2+2X_3}}\right)\right].$$

After some algebra it can be shown that $L_{KL}(\theta, \delta) = 2n\left[(1-\theta)log\left(\frac{1-\theta}{1-\delta}\right) + \theta log\left(\frac{\theta}{\delta}\right)\right]$, thus

$$\int_0^1 L_{KL}(\theta,\delta)\pi(\theta|X)d\theta = 2nE\left[(1-\theta)\ln\left(\frac{1-\theta}{1-\delta}\right) + \theta\ln\left(\frac{\theta}{\delta}\right)\bigg| X\right].$$

The goal is to minimize this expression with respect to $\delta$, which amounts to minimizing $-\ln(1-\delta)E[1-\theta|X] - \ln\delta E[\theta|X]$ because the reaming terms do not depend on $\delta$. Setting the first derivative with respect to $\delta$ to zero and checking the second order condition yields:

$$\frac{E[1-\theta|X]}{1-\delta} - \frac{E[\theta|X]}{\delta} = 0 \Rightarrow \delta = E[\theta|X].$$

Thus, as in the case of SEL, under KLL the Bayes estimator is also the posterior mean. Hence, from section 2.1.1 it follows that:

$$\hat{\theta}^{KLL} = E[\theta|X] = \frac{x_2 + 2x_3 + \alpha}{2n + \alpha + \beta} = \hat{\theta}^{SEL}.$$

The risk function of $\hat{\theta}^{KLL}$ is:

$$R(\theta,\hat{\theta}^{KLL}) = E_\theta[L_{KL}(\theta,\hat{\theta}^{KLL})] = 2nE_\theta\left[(1-\theta)\ln\left(\frac{1-\theta}{1-\hat{\theta}^{KLL}}\right) + \theta\ln\left(\frac{\theta}{\hat{\theta}^{KLL}}\right)\right]$$

$$= 2n\left[(1-\theta)(\ln(1-\theta) - E_\theta[\ln(1-\hat{\theta}^{KLL})]) + \theta(\ln\theta - E_\theta[\ln\hat{\theta}^{KLL}])\right].$$

This involves evaluating $E_\theta[\ln(1-\hat{\theta}^{KLL})]$ and $E_\theta[\ln\hat{\theta}^{KLL}]$. Consider $E_\theta[\ln\hat{\theta}^{KLL}] = E_\theta[\ln(X_2 + 2X_3 + \alpha)] - \ln(2n + \alpha + \beta)$. To simplify the problem recall that this is equivalent to $E_\theta[\ln(Y_1 + \alpha)] - \ln(2n + \alpha + \beta)$ where $Y_1$ is a Binomial$(2n,\theta)$ random variable; however, this expectation does not have a closed form. Similarly, by using the fact that $Y_1 + Y_2 = 2n$, it can be found that the evaluation of $E_\theta[\ln(1-\hat{\theta}^{KLL})]$ involves finding $E_\theta[\ln(Y_2 + \beta)]$ which has no closed form solution either.

2.1.3 Quadratic error loss

This loss can be seen as a weighted version of SEL and it has the following form: $w(\theta)(\delta - \theta)^2$, $w(\theta) > 0$, $\forall\, \theta \in \Theta$. Let $w(\theta) = [\theta(1-\theta)]^{-1}$. This form of $w(\theta)$ was chosen for mathematical convenience as it will become clear in the derivation of the decision rule. Thus, the loss function has the form: $L(\theta, \delta) = \frac{(\theta-\delta)^2}{\theta(1-\theta)}$. Under this kind of loss, the Bayes estimator is the mean of the distribution $w(\theta)\pi(\theta|X)$ (Lehman and Casella, 1998).

$$w(\theta)\pi(\theta|X) \propto \frac{1}{\theta(1-\theta)} \theta^{x_2+2x_3+\alpha-1}(1-\theta)^{2x_1+x_2+\beta-1}$$

$$= \theta^{x_2+2x_3+\alpha-2}(1-\theta)^{2x_1+x_2+\beta-2}$$

This corresponds to a $\text{Beta}(x_2 + 2x_3 + \alpha - 1, 2x_1 + x_2 + \beta - 1)$ provided that: $x_2 + 2x_3 + \alpha - 1 > 0, 2x_1 + x_2 + \beta - 1 > 0$. In such case the estimator is simply the mean of that distribution, that is:

$$\hat{\theta}^{QEL} = \frac{x_2 + 2x_3 + \alpha - 1}{2(x_1 + x_2 + x_3) + \alpha + \beta - 2}$$

$$= \frac{x_2 + 2x_3 + \alpha - 1}{2n + \alpha + \beta - 2} \;(\because x_1 + x_2 + x_3 = n)$$

Now, the two cases $x_2 + 2x_3 + \alpha - 1 \leq 0$ and $2x_1 + x_2 + \beta - 1 \leq 0$ are analyzed. Notice that $x_2 + 2x_3 + \alpha - 1$ and $2x_1 + x_2 + \beta - 1$ cannot be simultaneously smaller than or equal to zero because it would imply that there are no observations. From first principles, the expression $\int_0^1 w(\theta)(\theta - \hat{\theta}^{QEL})^2 \pi(\theta|x) d\theta$ is required to be finite (Lehman and Casella, 1998). If $x_2 + 2x_3 + \alpha - 1 \leq 0$, it implies that $(X_2, X_3) = (0,0)$ and $\alpha \leq 1$. Under these conditions:

$$\int_0^1 w(\theta)(\theta - \hat{\theta}^{QEL})^2 \pi(\theta|x) d\theta \propto \int_0^1 (\theta - \hat{\theta}^{QEL})^2 \theta^{\alpha-2}(1-\theta)^{2x_1+\beta-2} d\theta$$

$$= \int_0^1 \theta^\alpha (1-\theta)^{2x_1+\beta-2} d\theta - 2\hat{\theta}^{QEL} \int_0^1 \theta^{\alpha-1}(1-\theta)^{2x_1+\beta-2} d\theta$$

$$+ \left(\hat{\theta}^{QEL}\right)^2 \int_0^1 \theta^{\alpha-2}(1-\theta)^{2x_1+\beta-2} d\theta.$$

The first two integrals are finite whereas the third integral is not finite unless $\hat{\theta}^{QEL} = 0$. If $2x_1 + x_2 + \beta - 1 \leq 0$ then $(X_1, X_2) = (0,0)$ and $\beta \leq 1$, then:

$$\int_0^1 w(\theta)(\theta - \hat{\theta}^{QEL})^2 \pi(\theta|\boldsymbol{x}) d\theta \propto \int_0^1 \left(1 - \theta - (1 - \hat{\theta}^{QEL})\right)^2 \theta^{2x_3+\alpha-2}(1-\theta)^{\beta-2} d\theta$$

$$= \int_0^1 \theta^{2x_3+\alpha-2}(1-\theta)^\beta d\theta - 2(1 - \hat{\theta}^{QEL}) \int_0^1 \theta^{2x_3+\alpha-2}(1-\theta)^{\beta-1} d\theta$$

$$+ (1 - \hat{\theta}^{QEL})^2 \int_0^1 \theta^{2x_3+\alpha-2}(1-\theta)^{\beta-2} d\theta.$$

The first two integrals are finite. For the third integral to be finite $\hat{\theta}^{QEL}$ must be equal to one.

In summary, under the given prior and QEL, the Bayes estimator is:

$$\hat{\theta}^{QEL} = \begin{cases} \dfrac{x_2 + 2x_3 + \alpha - 1}{2n + \alpha + \beta - 2}, & \text{if } x_2 + 2x_3 + \alpha - 1 > 0 \text{ and } 2x_1 + x_2 + \beta - 1 > 0 \\ 0, & \text{if } x_2 + 2x_3 + \alpha - 1 \leq 0 \\ 1, & \text{if } 2x_1 + x_2 + \beta - 1 \leq 0 \end{cases}$$

A common situation is $x_2 + 2x_3 + \alpha - 1 > 0$, $2x_1 + x_2 + \beta - 1 > 0$, and in that case:

$$R(\theta, \hat{\theta}^{QEL}) = E_\theta \left[ w(\theta)(\hat{\theta}^{QEL} - \theta)^2 \right] = E_\theta \left[ \frac{1}{\theta(1-\theta)} \left( \frac{X_2 + 2X_3 + \alpha - 1}{2n + \alpha + \beta - 2} - \theta \right)^2 \right]$$

$$= \frac{1}{\theta(1-\theta)} \left( Var_\theta \left[ \frac{X_2 + 2X_3 + \alpha - 1}{2n + \alpha + \beta - 2} \right] + \left( E_\theta \left[ \frac{X_2 + 2X_3 + \alpha - 1 - \theta(2n + \alpha + \beta - 2)}{2n + \alpha + \beta - 2} \right] \right)^2 \right).$$

Notice that: $Var_\theta\left[\frac{X_2+2X_3+\alpha-1}{2n+\alpha+\beta-2}\right] = \frac{1}{(2n+\alpha+\beta-2)^2} Var_\theta[X_2 + 2X_3]$, $Var_\theta[X_2 + 2X_3]$ was previously derived in section 2.1.1 and it is equal to $2n\theta(1-\theta)$. On the other hand, the procedure to simplify the second summand is very similar to the one used for $R(\theta, \hat\theta^{SEL})$, and the final result is $\frac{(-\theta(\alpha+\beta-2)+\alpha-1)^2}{(2n+\alpha+\beta-2)^2}$. Therefore the risk has the form:

$$R(\theta, \hat\theta^{QEL}) = \frac{2n}{(2n+\alpha+\beta-2)^2} + \frac{(-\theta(\alpha+\beta-2)+\alpha-1)^2}{\theta(1-\theta)(2n+\alpha+\beta-2)^2}.$$

When $x_2 + 2x_3 + \alpha - 1 \leq 0$, that is, allele A is not observed and $\alpha \leq 1$, the risk is:

$$R(\theta, \hat\theta^{QEL}) = \frac{(\theta-0)^2}{\theta(1-\theta)} = \frac{\theta}{1-\theta},$$

while when $2x_1 + x_2 + \beta - 1 \leq 0$ (allele B is not observed and $\beta \leq 1$) the risk is

$$R(\theta, \hat\theta^{QEL}) = \frac{(\theta-1)^2}{\theta(1-\theta)} = \frac{1-\theta}{\theta}.$$

2.2 Derivation of minimax rules

To derive minimax rules the following theorem was used (Lehman and Casella, 1998):

*Theorem 1* Let $\Lambda$ be a prior and $\delta_\Lambda$ a Bayes rule with respect to $\Lambda$ with Bayes risk satisfying $r(\Lambda, \delta_\Lambda) = \sup_{\theta \in \Theta} R(\theta, \delta_\Lambda)$. Then: *i)* $\delta_\Lambda$ is minimax and *ii)* $\Lambda$ is least favorable.

A corollary that follows from this theorem is that if $\delta$ is a Bayes decision rule with respect to a prior $\Lambda$ and it has constant (not depending on $\theta$) frequentist risk, then it is also minimax and $\Lambda$ is least favorable, that is, it causes the greatest average loss. Thus, the approach was the following. Once a Bayes estimator was derived, it was determined if there were values of the hyperparameters such that $R(\theta, \delta)$ was constant; therefore, using these particular values of the hyperparameters, the resulting estimator was minimax. Notice that for SEL, by choosing the Beta$\left(\alpha = \sqrt{\frac{n}{2}}, \beta = \sqrt{\frac{n}{2}}\right)$ prior, the risk function $R(\theta, \hat{\theta}^{SEL})$ is constant and takes the form:

$R(\theta, \hat{\theta}^{Minimax_1}) = \left(4(1+\sqrt{2n})^2\right)^{-1}$. Hence, a minimax estimator is:

$$\hat{\theta}^{Minimax_1} = \frac{x_2 + 2x_3 + \sqrt{\frac{n}{2}}}{\sqrt{2n}(\sqrt{2n}+1)}.$$

On the other hand, it is easy to notice that provided $x_2 + 2x_3 + \alpha - 1 > 0$, $2x_1 + x_2 + \beta - 1 > 0$, $\hat{\theta}^{QEL}$ have a constant risk for $\alpha = \beta = 1$, that is, under a uniform(0,1) prior. Then:

$$\hat{\theta}^{Minimax_2} = \frac{x_2 + 2x_3}{2n} \text{ and } R(\theta, \hat{\theta}^{Minimax_2}) = \frac{1}{2n} \; \forall \, \theta \in \Theta.$$

In the case of the Bayes estimator derived under KLL, the risk function involves the evaluation of a finite sum that does not have a closed form solution. Although an approximation based on the Taylor series expansion of $\ln(Y_1 + \alpha)$ and $\ln(Y_2 + \beta)$ could be found, it turns out that this function cannot be made independent of $\theta$ by manipulating the hyperparameters $\alpha$ and $\beta$. Consequently, theorem 1 could not be used here to find a minimax estimator. Because of this, hereinafter only SEL and QEL will be used to obtain Bayes and minimax decision rules.

2.3 Extension to k loci

When the interest is in estimating allelic frequencies at several loci, i.e., the parameter is vector-valued, it could seem natural to compute the real-valued estimators presented in sections 2.1 and 2.2 at each locus and combine them to obtain the desired estimator. The question is: Do these estimators preserve the properties of Bayesness and minimaxity of their univariate counterparts? In this section we show that this is the case under certain assumptions, and therefore, Bayes estimation of vector-valued parameters reduces to estimation of each of its components.

Let $\boldsymbol{\theta} = (\theta_1, \theta_2, \ldots, \theta_k)$ be the vector containing the frequencies of the "reference" alleles for k loci, $\boldsymbol{X} = (\boldsymbol{X}_1, \boldsymbol{X}_2, \ldots, \boldsymbol{X}_k)$ the vector containing the number of individuals for every genotype at every locus where $\boldsymbol{X}_i = (X_{1i}, X_{2i}, X_{3i})$, $i = 1, 2, \ldots, k$, and $\boldsymbol{\delta} = (\delta_1, \delta_2, \ldots, \delta_k)$ a vector-valued estimator of $\boldsymbol{\theta}$. Consider a general additive loss function of the form: $L(\boldsymbol{\theta}, \boldsymbol{\delta}(\boldsymbol{X})) = \sum_{i=1}^{k} L(\theta_i, \delta_i(\boldsymbol{X}))$. Assuming linkage equilibrium we have $\pi(\boldsymbol{X}|\boldsymbol{\theta}) = \prod_{i=1}^{k} \pi(\boldsymbol{X}_i|\theta_i)$ and using independent priors it follows that $\pi(\boldsymbol{\theta}|\boldsymbol{X}) = \prod_{i=1}^{k} \pi(\theta_i|\boldsymbol{X}_i)$. To obtain a Bayes estimator, the following expression has to be minimized with respect to $\delta_i, \forall i = 1, 2, \ldots, k$:

$$\int_{\Theta_1} \cdots \int_{\Theta_k} L(\boldsymbol{\theta}, \boldsymbol{\delta}(\boldsymbol{X})) \pi(\boldsymbol{\theta}|\boldsymbol{X}) d\theta_1 \cdots d\theta_k = \int_{\Theta_1} \cdots \int_{\Theta_k} \left(\sum_{i=1}^{k} L(\theta_i, \delta_i(\boldsymbol{X}))\right) \pi(\boldsymbol{\theta}|\boldsymbol{X}) d\theta_1 \cdots d\theta_k$$

$$= \sum_{i=1}^{k} \int_{\Theta_1} \cdots \int_{\Theta_k} L(\theta_i, \delta_i(\boldsymbol{X})) \prod_{j=1}^{k} \pi(\theta_j|\boldsymbol{X}_j) d\theta_1 \cdots d\theta_k$$

the $h^{th}$ integral in the summation ($h = 1, 2, \ldots, k$) can be written as:

$$\int_{\Theta_h} L(\theta_h, \delta_h(\boldsymbol{X})) \pi(\theta_h|\boldsymbol{X}_h) d\theta_h \int_{\Theta_1} \cdots \int_{\Theta_{h-1}} \int_{\Theta_{h+1}} \cdots \int_{\Theta_k} \prod_{j \neq h} \pi(\theta_j|\boldsymbol{X}_j) d\theta_1 \cdots d\theta_{h-1} d\theta_{h+1} \cdots d\theta_k$$

$$= \int_{\Theta_h} L(\theta_h, \delta_h) \pi(\theta_h|\boldsymbol{X}_h) d\theta_h.$$

From the result above, it follows that Bayes estimation of $\boldsymbol{\theta}$ reduces to that of its components. Therefore, under linkage equilibrium, independent priors and an additive loss it follows that $\widehat{\boldsymbol{\theta}}^{Bayes} = (\hat{\theta}_1^{Bayes}, \hat{\theta}_2^{Bayes}, \ldots, \hat{\theta}_k^{Bayes})$. Applying the results derived previously, a minimax estimator is the vector $\widehat{\boldsymbol{\theta}}^{Minimax_1} \in \mathbb{R}^k$, whose $i^{th}$ entry is $\frac{x_{2i} + 2x_{3i} + \sqrt{\frac{n}{2}}}{\sqrt{2n}(\sqrt{2n}+1)}$. Another minimax estimator of $\boldsymbol{\theta}$ is obtained by posing k independent uniform(0,1) priors and the $i^{th}$ element of $\widehat{\boldsymbol{\theta}}^{Minimax_2} \in \mathbb{R}^k$ has the form $\frac{x_{2_i} + 2x_{3_i}}{2n}$, provided $x_{2_i} + 2x_{3_i} + \alpha - 1 > 0$ and $2x_{1_i} + x_{2_i} + \beta - 1 > 0 \ \forall \ i = 1, 2, \ldots, k$.

## 2.4 Multiallelic loci

In this section, the general case of two or more alleles per locus is discussed. The approach is the same used in the biallelic loci case. In first place, an arbitrary locus $i$ having $n_i$ alleles is considered, and then the results are expanded to the multiple loci scenario. Let $\theta_{1_i}, \theta_{2_i}, \ldots, \theta_{n_i}$ be the frequencies of the $n_i$ alleles of locus $i$ and $X_{1_i}, X_{2_i}, \ldots, X_{N_i}$ random variables indicating the number of individuals having each one of the $N_i$ possible genotypes formed from the $n_i$ different alleles, $i = 1, 2, \ldots, k$. Notice that for diploid organisms $N_i = \binom{n_i}{2} + n_i$. The sampling model can be written as a multinomial distribution of dimension $N_i$; however, as discussed previously, an equivalent sampling model in terms of the counts for every allelic variant can be used. This approach is simpler because $N_i$ could be large. Hence, let $Y_{1_i}, Y_{2_i}, \ldots, Y_{n_i}$ be random variables indicating the counts of each one of the $n_i$ allelic variants at locus $i$: $A_{1_i}, A_{2_i}, \ldots, A_{n_i}; i = 1, 2, \ldots, k$. A multinomial distribution with parameters $\boldsymbol{\theta}_i = (\theta_{1_i}, \theta_{2_i}, \ldots, \theta_{n_i})$ and $2n$ is assigned to $\boldsymbol{Y}_i = (Y_{1_i}, Y_{2_i}, \ldots, Y_{n_i})$. The parametric space is denoted by $\Theta$ and corresponds to $[0,1] \times [0,1] \times \cdots \times [0,1]$, an $n_i$-dimensional unit hypercube. The prior assigned to $\boldsymbol{\theta}_i$ is a Dirichlet distribution with hyperparameters $\boldsymbol{\alpha}_i = (\alpha_{1_i}, \alpha_{2_i}, \ldots, \alpha_{n_i})$. With this setting, conjugacy holds and therefore the posterior is a Dirichlet $(\alpha_{1_i} + y_{1_i}, \alpha_{2_i} + y_{2_i}, \ldots, \alpha_{n_i} + y_{n_i})$. Under an additive SEL of the form $\sum_{j_i=1}^{n_i} (\hat{\theta}_{j_i} - \theta_{j_i})^2$ the Bayes estimator of $\boldsymbol{\theta}_i$ is given by the vector of posterior means, that is:

$$\widehat{\boldsymbol{\theta}}_i^{M-SEL} = (\hat{\theta}_{j_i})_{n_i \times 1} = \frac{\alpha_{j_i} + Y_{j_i}}{2n + \sum_{j_i=1}^{n_i} \alpha_{j_i}},$$

where the "M" in the super-index stands for multiple loci. The risk of this estimator is:

$$R(\boldsymbol{\theta}_i, \widehat{\boldsymbol{\theta}}_i^{M-SEL}) = E_{\boldsymbol{\theta}_i}\left[\sum_{j_i=1}^{n_i} (\widehat{\boldsymbol{\theta}}_{j_i}^{M-SEL} - \theta_{j_i})^2\right],$$

that can be shown to have the form:

$$\sum_{j_i=1}^{n_i} \frac{\theta_{j_i}^2 \left(\left(\sum_{l_i=1}^{n_i} \alpha_{l_i}\right)^2 - 2n\right) + \theta_{j_i}\left(2n - 2\alpha_{j_i}\sum_{l_i=1}^{n_i}\alpha_{l_i}\right) + \alpha_{j_i}^2}{\left(2n + \sum_{l_i=1}^{n_i}\alpha_{l_i}\right)^2}.$$

To find a minimax estimator, theorem 1 is invoked again. Based on the results from the biallelic case, intuition suggests trying the following values for the hyperparameters: $\alpha_{j_i} = \sqrt{2n}/n_i$, $\forall\, j_i = 1,2,\ldots,n_i$. Then, after simplification:

$$R(\boldsymbol{\theta}_i, \widehat{\boldsymbol{\theta}}_i^{M-Minimax_1}) = \frac{\left(\frac{n_i-2}{n_i}\right)\sum_{j_i=1}^{n_i}\theta_{j_i} + \frac{1}{n_i}}{(\sqrt{2n}+1)^2} = \frac{\frac{n_i-1}{n_i}}{(\sqrt{2n}+1)^2},$$

where the last equality follows from the fact that $\sum_{j_i=1}^{n_i}\theta_{j_i} = 1$. Hence, under these particular values of the hyperparameters, the risk is constant and therefore, a minimax estimator is:

$$\widehat{\boldsymbol{\theta}}_i^{M-Minimax_1} = (\hat{\theta}_{j_i})_{n_i \times 1} = \frac{y_{j_i} + \frac{\sqrt{2n}}{n_i}}{\sqrt{2n}(\sqrt{2n}+1)}.$$

Now consider an additive loss of the form $\sum_{j_i=1}^{n_i} w(\theta_{j_i})(\hat{\theta}_{j_i} - \theta_{j_i})^2$, $w(\theta_{j_i}) > 0\ \forall\ \theta_{j_i} \in \Theta$. Again, $w(\theta_{j_i})$ is chosen for convenience and it is defined as $w(\theta_{j_i}) = \theta_{j_i}^{-1}\ \forall\ j_i = 1,2,\ldots,n_i$. In this case the function to be minimized is:

$$\int_\Theta \sum_{j_i=1}^{n_i} w(\theta_{j_i})(\hat{\theta}_{j_i} - \theta_{j_i})^2 \pi(\boldsymbol{\theta}_i|\boldsymbol{Y}_i)\,d\theta_i = \sum_{j_i=1}^{n_i}\int_\Theta w(\theta_{j_i})(\hat{\theta}_{j_i} - \theta_{j_i})^2 \pi(\boldsymbol{\theta}_i|\boldsymbol{Y}_i)d\theta_i$$

which is equivalent to minimizing every term in the summation. Therefore, for every term this is the same problem discussed in the biallelic case, and it follows that for $j_i = 1, 2, \ldots, n_i$, $\hat{\theta}_{j_i}$ is the expectation of $\theta_{j_i}$ taken with respect to the density $v(\theta) = \frac{w(\theta_{j_i})\pi(\theta_i|Y_i)}{\int_\Theta w(\theta_{j_i})\pi(\theta_i|Y_i)d\theta_i}$ provided $\int_\Theta w(\theta_{j_i})\pi(\theta_i|Y_i)d\theta_i < \infty$ (Lehmann and Casella, 1998). Thus,

$$w(\theta_{j_i})\pi(\theta_i|Y_i) \propto \theta_{1_i}^{\alpha_{1_i}+y_{1_i}-1} \cdots \theta_{(j-1)_i}^{\alpha_{(j-1)_i}+y_{(j-1)_i}-1} \theta_{j_i}^{\alpha_{j_i}+y_{j_i}-2} \theta_{(j+1)_i}^{\alpha_{(j+1)_i}+y_{(j+1)_i}-1} \cdots \theta_{n_i}^{\alpha_{n_i}+y_{n_i}-1}.$$

This is the kernel of a Dirichlet$(\alpha_{1_i} + y_{1_i}, \ldots, \alpha_{j_i} + y_{j_i} - 1, \ldots, \alpha_{n_i} + y_{n_i})$ density provided $\alpha_{j_i} + y_{j_i} - 1 > 0$. In this case, $\hat{\theta}_{j_i} = \frac{\alpha_{j_i}+y_{j_i}-1}{\sum_{j_i=1}^{n_i}\alpha_{j_i}+2n-1}$ $\forall j_i = 1, 2, \ldots, n_i$. If $\alpha_{j_i} + y_{j_i} - 1 \leq 0$, it must be that $y_{j_i} = 0, \alpha_{j_i} \leq 1$ and following the same reasoning used for biallelic loci, it turns out that the estimator is $\hat{\theta}_{j_i} = 0$. In summary, under this additive quadratic loss function, for $j_i = 1, 2, \ldots, n_i$, the Bayes estimator under the Dirichlet prior and the given loss function is:

$$\hat{\boldsymbol{\theta}}_i^{M-QEL} = \left(\hat{\theta}_{j_i}^{M-QEL}\right)_{n_i \times 1} = \begin{cases} \frac{\alpha_{j_i}+y_{j_i}-1}{\sum_{j_i=1}^{n_i}\alpha_{j_i}+2n-1}, & \text{if } \alpha_{j_i}+y_{j_i}-1 > 0 \\ 0, & \text{if } \alpha_{j_i}+y_{j_i}-1 \leq 0 \end{cases}$$

The risk of this estimator when $\alpha_{j_i} + y_{j_i} - 1 > 0 \,\forall j_i = 1, 2, \ldots, n_i$ is:

$$R(\boldsymbol{\theta}_i, \hat{\boldsymbol{\theta}}_i^{M-QEL}) = \sum_{j_i=1}^{n_i} E_\theta \left[ w(\theta_{j_i})\left(\hat{\theta}_{j_i}^{M-QEL} - \theta_{j_i}\right)^2 \right]$$

The derivation is similar to the one in the biallelic case and $R(\boldsymbol{\theta}_i, \hat{\boldsymbol{\theta}}_i^{M-QEL})$ has the form:

$$\frac{2n(n_i - 1) + \sum_{j_i=1}^{n_i}\frac{(\alpha_{j_i}-1)^2}{\theta_{j_i}} + \left(\sum_{j_i=1}^{n_i}\alpha_{j_i}-1\right)\left(\left(\sum_{j_i=1}^{n_i}\alpha_{j_i}-1\right) - 2\sum_{j_i=1}^{n_i}(\alpha_{j_i}-1)\right)}{\left(\sum_{j_i=1}^{n_i}\alpha_{j_i}+2n-1\right)^2}.$$

In the light of theorem 1, it is easy to see that provided $\alpha_{j_i} + y_{j_i} - 1 > 0 \,\forall j_i = 1, 2, \ldots, n_i$, by assigning a Dirichlet prior with all hyperparameters equal to one, the risk is constant and equal to $\frac{(n_i-1)}{2n+n_i-1}$. Consequently, the minimax estimator obtained here is $\widehat{\boldsymbol{\theta}}_i^{M-Minimax_2} \in \mathbb{R}^{n_i}$ whose $j^{th}$ entry is $\frac{y_{j_i}}{2n+n_i-1}$. Details of the derivations of $\widehat{\boldsymbol{\theta}}_i^{M-QEL}$ and the risk functions are presented in the appendix.

Under the assumption of linkage equilibrium, posing independent priors and considering an additive loss function, the extension to k loci is straightforward and it is basically the same for the biallelic loci scenario. The parameter is $\boldsymbol{\theta} = (\boldsymbol{\theta}_1, \boldsymbol{\theta}_2, \dots, \boldsymbol{\theta}_k)$ where $\boldsymbol{\theta}_i \in \mathbb{R}^{n_i}$ contains the frequencies of each allele in locus $i, i = 1,2,\dots,k$. Hence, independent Dirichlet($\boldsymbol{\alpha}_i$) priors are assigned to the elements of $\boldsymbol{\theta}$. Let $\boldsymbol{Y} = (\boldsymbol{Y}_1, \boldsymbol{Y}_2, \dots, \boldsymbol{Y}_k)$ be a vector containing the counts of each allele at each loci, that is, $\boldsymbol{Y}_i \in \mathbb{R}^{n_i}$ contains the counts of the $n_i$ alleles in locus $i$. The loss function has the form $L(\boldsymbol{\theta}, \delta(\boldsymbol{Y})) = \sum_{i=1}^{k} L(\boldsymbol{\theta}_i, \delta_i(\boldsymbol{Y}))$. Then, as in the biallelic case, the key property $\pi(\boldsymbol{\theta}|\boldsymbol{Y}) = \prod_{i=1}^{k} \pi(\boldsymbol{\theta}_i|\boldsymbol{Y}_i)$ holds and therefore, finding decision rules to estimate $\boldsymbol{\theta}$ amounts to finding decision rules to estimate its components: $\boldsymbol{\theta}_1, \boldsymbol{\theta}_2, \dots, \boldsymbol{\theta}_k$. In this case, the estimators are denoted as $\widehat{\boldsymbol{\theta}}^{M-SEL}, \widehat{\boldsymbol{\theta}}^{M-QEL}, \widehat{\boldsymbol{\theta}}^{M-Minimax_1}$ and $\widehat{\boldsymbol{\theta}}^{M-Minimax_2}$.

Admissibility of one-dimensional and vector-valued estimators was established using a theorem found in Lehmann and Casella (1998) which is restated for the reader's convenience.

*Theorem 2* For a possibly vector-valued parameter $\boldsymbol{\theta}$, suppose that $\delta^\pi$ is a Bayes estimator having finite Bayes risk with respect to a prior density $\pi$ which is positive for all $\boldsymbol{\theta} \in \Theta$, and that the risk function of every estimator $\delta$ is a continuous function of $\boldsymbol{\theta}$. Then $\delta^\pi$ is admissible.

A key condition of this theorem is the continuity of the risk for all decision rules. For exponential families, this condition holds (Lehmann and Casella, 1998) and given that all distributions considered here are exponential families, the condition is met.

## 3. Results

For biallelic loci, the Bayesian decision rules derived under SEL and KLL were found to be the same. Notice that this estimator can be rewritten as $\frac{x_2+2x_3}{2n}\left(\frac{2n}{2n+\alpha+\beta}\right)+\frac{\alpha}{\alpha+\beta}\left(\frac{\alpha+\beta}{2n+\alpha+\beta}\right)$, which is a convex combination of the maximum likelihood estimator (MLE) and the prior mean. On the other hand, the Bayesian decision rule found under the QEL depends on the values taken by $x_2 + 2x_3 + \alpha - 1$ and $2x_1 + x_2 + \beta - 1$. As discussed previously, when at least one observation is done (at least one genotyped individual) these quantities cannot be simultaneously smaller or equal than zero, since it $\alpha > 0, \beta > 0$ and in case of observing one or more genotypes, at least one of the random variables $X_1, X_2$ and $X_3$ would take a value greater or equal than one. Notice that when $x_2 + 2x_3 > 0$ and $2x_1 + x_2 > 0$, $\hat{\theta}^{Minimax_2}$ does exist and it is equivalent to the MLE. Thus, it has been shown that the MLE is also minimax and that the uniform(0,1) prior is least favorable for estimating $\theta$ under QEL. Moreover, a Beta$\left(\sqrt{\frac{n}{2}}, \sqrt{\frac{n}{2}}\right)$ prior was also found to be least favorable under SEL. When $x_2 + 2x_3 + \alpha - 1 > 0$, $2x_1 + x_2 + \beta - 1 > 0$, the estimator $\hat{\theta}^{QEL}$ can be rewritten as $\frac{x_2+2x_3}{2n}\left(\frac{2n}{2n+\alpha+\beta-2}\right)+\frac{\alpha}{\alpha+\beta}\left(\frac{\alpha+\beta}{2n+\alpha+\beta-2}\right)+\frac{1}{2}\left(\frac{-2}{2n+\alpha+\beta-2}\right)$ a linear combination of the MLE, the prior mean and the scalar $1/2$. Regarding admissibility of the one-dimensional estimators, $\hat{\theta}^{SEL}, \hat{\theta}^{Minimax_1}$ and $\hat{\theta}^{Minimax_2}$ have finite Bayesian risks and therefore, by theorem 2, they are admissible. For $\hat{\theta}^{QEL}$ the property holds provided $\alpha > 1, \beta > 1$. For the case of k loci, under additive loss functions, the risks are additive and therefore the Bayes risks too. Hence, the estimators $\widehat{\boldsymbol{\theta}}^{SEL}, \widehat{\boldsymbol{\theta}}^{Minimax_1}$ and $\widehat{\boldsymbol{\theta}}^{Minimax_2}$ are admissible, and if $\alpha_i > 1, \beta_i > 1 \; \forall \; i = 1,2, \ldots, k$, $\widehat{\boldsymbol{\theta}}^{QEL}$ is also admissible.

In the multiallelic case, notice that $\widehat{\boldsymbol{\theta}}_i^{M-Minimax_1}$ reduces to its biallelic version ($n_i = 2$) because $y_{j_i} = x_{2_i} + 2x_{3_i}$. This happens because $\widehat{\boldsymbol{\theta}}_i^{M-Minimax_1}$ was derived from a Bayes estimator under SEL; however, when $n_i = 2$, $\widehat{\boldsymbol{\theta}}_i^{M-Minimax_2}$ does not reduce to $\hat{\theta}_i^{Minimax_2}$, but the estimators only differ in the denominator which is $2n + 1$ for $\widehat{\boldsymbol{\theta}}_i^{M-Minimax_2}$ and $2n$ for $\hat{\theta}_i^{Minimax_2}$; hence, for large $n$ the estimators are very close. These results for the one locus case also hold for the case of several loci given the way in which the multiple-loci estimators were derived. Regarding admissibility in the multiallelic setting, for the single-locus case, in the light of theorem 2 $\widehat{\boldsymbol{\theta}}_i^{M-SEL}, \widehat{\boldsymbol{\theta}}_i^{M-Minimax_1}$ and $\widehat{\boldsymbol{\theta}}_i^{M-Minimax_2}$ are admissible and provided $\alpha_{j_i} > 1, \forall j_i = 1,2, \dots, n_i$, $\widehat{\boldsymbol{\theta}}_i^{M-QEL}$ is also admissible. The same reasoning used in the biallelic case shows that for k loci and $n_i$ alleles per locus, $\widehat{\boldsymbol{\theta}}^{M-SEL}, \widehat{\boldsymbol{\theta}}^{M-Minimax_1}$ and $\widehat{\boldsymbol{\theta}}^{M-Minimax_2}$ are admissible, as well as $\widehat{\boldsymbol{\theta}}^{M-QEL}$ when $\alpha_{j_i} > 1, \forall j_i = 1,2, \dots, n_i, \forall i = 1,2, \dots, k$.

3.1 Comparison of estimators

Because of the interest in addressing situations in which the proposed estimators may differ substantially from each other, in this section they are compared by finding general algebraic expressions that help in analyzing how they differ. These comparisons are basically related to values of the hyperparameters, to the allelic counts and to sample size.

The risks of the estimators here cannot be compared directly because their corresponding loss functions measure the distance between estimators and estimands in different ways. Consequently, the precision of the estimators was compared using their frequentist (conditional on $\theta$) variances. It is enough to carry out comparisons for an arbitrary locus for the biallelic and multiallelic cases.

The magnitudes of all point estimators were compared with the MLE and against each other by finding their ratios. In each case, a short interpretation of the resulting expression is done in order to provide some settings under which the estimators show considerable differences. For the biallelic case, the ratio of estimator Z and the MLE is defined as $\delta^Z$. Thus, after simplification:

$$\delta^{SEL} = \frac{2n}{2n + \alpha + \beta}\left(1 + \frac{\alpha}{x_2 + 2x_3}\right),$$

$$\delta^{SEL} > (<)1 \Leftrightarrow x_2 + 2x_3 < (>) \frac{2n\alpha}{\alpha + \beta}.$$

Thus, given $n$, $(x_2, x_3)$ and $\alpha$, the ratio is larger as $\beta \downarrow 0$ and decreases monotonically as $\beta \to \infty$. On the other hand, if $\alpha \downarrow 0$ and $\beta \to \infty$ the ratio is smaller than one for fixed $n$. For very low counts of AA and AB genotypes, i.e., small $x_2$ and $x_3$, and $\alpha$ not close to zero, the ratio tends to be greater than one.

Recall that $\hat{\theta}^{QEL}$ depends on $x_2 + 2x_3 + \alpha - 1$ and $2x_1 + x_2 + \beta - 1$. When $x_2 + 2x_3 + \alpha - 1 > 0$, $2x_1 + x_2 + \beta - 1 > 0$, it follows that:

$$\delta^{QEL} = \frac{2n}{2n + \alpha + \beta - 2}\left(1 + \frac{\alpha - 1}{x_2 + 2x_3}\right),$$

$$\delta^{QEL} > (<)1 \Leftrightarrow x_2 + 2x_3 < (>) \frac{2n(\alpha - 1)}{\alpha + \beta - 2}.$$

Notice that if $\alpha < 1$, (which requires $x_2 + 2x_3 \geq 1$) and $\beta > 2 - \alpha$ or $\alpha > 1$, and $\beta < 2 - \alpha$, then the ratio is always smaller than one and the difference between estimators increases as genotypes AB and BB are more frequent, i.e., large $x_2$ and $x_3$. Moreover, when $\alpha > 1$, and $\beta > 2 - \alpha$, if $\alpha \downarrow 1$ and $\beta \to \infty$ the ratio will also be smaller than one. Similar interpretations can be done for the case of the ratio being greater than one. Recall that when $x_2 + 2x_3 + \alpha - 1 \leq 0$ or $2x_1 + x_2 + \beta - 1 \leq 0$, $\hat{\theta}^{QEL}$ matches the MLE, and when both alleles are observed, $\hat{\theta}^{Minimax_2}$ matches the MLE. Figure 1 shows the behavior of $\delta^{SEL}$ and $\delta^{QEL}$ as a function of the hyperparameter $\beta$ in two scenarios. Under scenario 1 the genotype counts are $(x_2, x_3) = (10, 25)$, while in scenario 2 $(x_2, x_3) = (250, 313)$. In each case, two sample sizes are considered: 1382 and 691.

For $\hat{\theta}^{Minimax_1}$:

$$\delta^{Minimax_1} = \frac{\sqrt{2n}\left(x_2 + 2x_3 + \sqrt{\frac{n}{2}}\right)}{(\sqrt{2n} + 1)(x_2 + 2x_3)},$$

$$\delta^{Minimax_1} > 1 \Leftrightarrow x_2 + 2x_3 < n,$$

since $\frac{x_2 + 2x_3}{n} := \hat{p} \in [0, 1]$ is the observed frequency (also the MLE) of the reference allele B, it follows that: $x_2 + 2x_3 = n\hat{p} < n$ if and only if $\hat{p} < 1$. The same rationale shows that $\delta^{Minimax_1}$ is never smaller than one, i.e., $\hat{\theta}^{Minimax_1}$ is never smaller than the MLE. For given $n$, when allele B is very rare, i.e. $(x_2, x_3) \to (0,0)$ the ratio tends to infinite. Figure 2 shows the behavior of $\delta^{Minimax_1}$ as a function of the observed frequency of the reference allele B for four different sample sizes (200, 800, 2000, and 10000).

For the case $x_2 + 2x_3 + \alpha - 1 > 0$, $2x_1 + x_2 + \beta - 1 > 0$, the Bayes estimator $\hat{\theta}^{QEL}$ differs from $\hat{\theta}^{SEL}$ in that the numerator of $\hat{\theta}^{QEL}$ is equal to the numerator of $\hat{\theta}^{SEL}$ minus one and its denominator is equal to the denominator of $\hat{\theta}^{SEL}$ minus two. Consequently, for moderate and large $n$, the estimators are very similar. Thus, only $\hat{\theta}^{SEL}$ is compared with $\hat{\theta}^{Minimax_1}$ here.

$$\frac{\hat{\theta}^{SEL}}{\hat{\theta}^{Minimax_1}} = \frac{(x_2 + 2x_3 + \alpha)(2n + \sqrt{2n})}{\left(x_2 + 2x_3 + \sqrt{\frac{n}{2}}\right)(2n + \alpha + \beta)}$$

$$\frac{\hat{\theta}^{SEL}}{\hat{\theta}^{Minimax_1}} > (<)1 \Leftrightarrow \frac{x_2 + 2x_3 + \alpha}{x_2 + 2x_3 + \sqrt{\frac{n}{2}}} > (<) \frac{2n + \alpha + \beta}{2n + \sqrt{2n}}.$$

For instance, if $\alpha > \sqrt{\frac{n}{2}}$ and $\sqrt{2n} > \alpha + \beta$ which implies $\sqrt{n}\left(\sqrt{2} - \frac{1}{\sqrt{2}}\right) > \beta$, the ratio is greater than one. The behavior of this ratio as a function of $\beta$ is also shown in Figure 1.

The procedure is analogous for the case of multiple alleles. Define the ratio of estimator Z to the MLE as $\gamma^Z$. An arbitrary locus $i$ and an allele $j$ are considered.

$$\gamma^{M-SEL}{ji} = \frac{2n(\alpha_{j_i} + y_{j_i})}{y_{j_i}(2n + \alpha)}, \alpha^* = \sum_{k_i=1}^{n_i} \alpha_{k_i}$$

$$\gamma^{M-SEL}{ji} > (<) 1 \Leftrightarrow \alpha_{j_i} > (<) \frac{y_{j_i}\alpha^*}{2n}.$$

For example, when allele $j$ is not observed, the ratio is always greater than one. For a given $\alpha_{j_i}$, the ratio increases as $n$ increases but the count of the allele remains constant or has a very small increase as in the case of a rare allelic variant.

On the other hand:

$$\gamma^{M-QEL}{ji} = \frac{2n(\alpha_{j_i} + y_{j_i} - 1)}{y_{j_i}(2n + \alpha - 1)}$$

$$\gamma^{M-QEL}{ji} > (<) 1 \Leftrightarrow \alpha_{j_i} - 1 > (<) \frac{y_{j_i}(\alpha - 1)}{2n}.$$

For $\alpha_{j_i} \gg 1$, large $n$ and low frequency of allele $j$ the ratio will be greater than one. On the other hand, if $\alpha_{j_i} < 1$ and $\alpha^* > 1$, then the ratio will be smaller than one disregarding of $y_{j_i}$ and the sample size.

For $\left(\widehat{\theta}_i^{M-Minimax_1}\right)_j$:

$$\gamma^{M-Minimax_1}{ji} = \frac{\sqrt{2n}\left(y_{j_i} + \frac{\sqrt{2n}}{n_i}\right)}{y_{j_i}(\sqrt{2n} + 1)}$$

$$\gamma^{M-Minimax_1 ji} > (<) 1 \Leftrightarrow y_{j_i} < (>) \frac{2n}{n_i},$$

where $n_i$ is the number of alleles at locus $i$. If allele $j$ is not observed ($y_{j_i} = 0$) then the ratio is always greater than one. If allele $j$ is fixed then $y_{j_i} = 2n$ and the ratio is always smaller than one and the larger the number of alelles at locus $i$, the larger the difference between the MLE and $\left(\widehat{\boldsymbol{\theta}}_i^{M-Minimax_1}\right)_j$.

In the multiallelic case, the estimator $\widehat{\boldsymbol{\theta}}_i^{M-Minimax_2}$ is not equal to the MLE; therefore, the ratio of $\widehat{\boldsymbol{\theta}}_i^{M-Minimax_2}$ and the MLE has to be computed.

$$\gamma^{M-Minimax_2 ji} = \frac{2n}{2n - n_i - 1} > 1 \ \forall \ n \geq 1, \forall \ n_i \geq 2,$$

consequently, $\left(\widehat{\boldsymbol{\theta}}_i^{M-Minimax_2}\right)_j$ is always larger than the MLE and the difference increases as the number of allelic variants at locus $i$ increases.

Again, as in the biallelic case, the differences between $\left(\widehat{\boldsymbol{\theta}}_i^{M-SEL}\right)_j$ and $\left(\widehat{\boldsymbol{\theta}}_i^{M-QEL}\right)_j$ are negligible and therefore this ratio is not computed and only $\left(\widehat{\boldsymbol{\theta}}_i^{M-SEL}\right)_j$ is compared with $\left(\widehat{\boldsymbol{\theta}}_i^{M-Minimax_1}\right)_j$ and $\left(\widehat{\boldsymbol{\theta}}_i^{M-Minimax_2}\right)_j$.

$$\frac{\left(\widehat{\boldsymbol{\theta}}_i^{M-SEL}\right)_j}{\left(\widehat{\boldsymbol{\theta}}_i^{M-Minimax_1}\right)_j} = \frac{\left(y_{j_i} + \alpha_{j_i}\right)\left(2n + \sqrt{2n}\right)}{\left(y_{j_i} + \frac{\sqrt{2n}}{n_i}\right)(2n + \alpha^*)}$$

$$\frac{\left(\widehat{\boldsymbol{\theta}}_i^{M-SEL}\right)_j}{\left(\widehat{\boldsymbol{\theta}}_i^{M-Minimax_1}\right)_j} > (<) 1 \Leftrightarrow \frac{y_{j_i} + \alpha_{j_i}}{y_{j_i} + \frac{\sqrt{2n}}{n_i}} > (<) \frac{2n + \alpha^*}{2n + \sqrt{2n}}$$

This case is similar to the biallelic case. When $\alpha_{j_i} > \frac{\sqrt{2n}}{n_i}$ and $\sqrt{2n} > \alpha^*$ which implies $\alpha_{j_i} > \frac{\alpha^*}{n_i}$, the ratio is bigger than one, and for fixed $n$ it increases as $\alpha_{j_i}$ increases and/or the number of alleles at locus $i$ increases. On the other hand, for fixed number of allelic variants, fixed $n$ and fixed $\alpha_{j_i}$, the ratio decreases as $\alpha^*$ increases.

$$\frac{\left(\hat{\boldsymbol{\theta}}_i^{M-SEL}\right)_j}{\left(\hat{\boldsymbol{\theta}}_i^{M-Minimax_2}\right)_j} = \frac{(2n - n_i - 1)(y_{j_i} + \alpha_{j_i})}{(2n + \alpha^*)y_{j_i}}$$

$$\frac{\left(\hat{\boldsymbol{\theta}}_i^{M-SEL}\right)_j}{\left(\hat{\boldsymbol{\theta}}_i^{M-Minimax_2}\right)_j} > (<)1 \Leftrightarrow \frac{2n - n_i - 1}{2n + \alpha^*} > (<)\frac{y_{j_i}}{y_{j_i} + \alpha_{j_i}}.$$

Now consider the frequentist variances for an arbitrary locus in the biallelic case:

$$Var_\theta[\hat{\theta}^{SEL}] = \frac{2n\theta(1-\theta)}{(2n + \alpha + \beta)^2}$$

$$Var_\theta[\hat{\theta}^{Minimax_1}] = \frac{\theta(1-\theta)}{(\sqrt{2n}+1)^2}$$

$$Var_\theta[\hat{\theta}^{Minimax_2}] = Var_\theta[\hat{\theta}^{ML}] = \frac{\theta(1-\theta)}{2n}$$

If $x_2 + 2x_3 + \alpha - 1 > 0$, $2x_1 + x_2 + \beta - 1 > 0$, then:

$$Var_\theta[\hat{\theta}^{QEL}] = \frac{2n\theta(1-\theta)}{(2n + \alpha + \beta - 2)^2}$$

Because the hyperparameters $\alpha$ and $\beta$ are positive and $n \geq 1$, the variances of $\hat{\theta}^{SEL}$ and $\hat{\theta}^{Minimax_1}$ are uniformly smaller than the variance of the conventional estimator, the MLE, except at the boundaries of the parameter space where all of them are zero. If $\alpha + \beta > 2$ then the variance of $\hat{\theta}^{QEL}$ is also uniformly smaller than the variance of the MLE, provided both alleles are observed. For $\hat{\theta}^{SEL}$ and $\hat{\theta}^{QEL}$ the differences increase as the hyperparameters increase while for $\hat{\theta}^{Minimax_1}$ the difference depends entirely on $n$. Given $\alpha$ and $\beta$, as the sample size tends to infinite, all variance ratios tend to one. In addition, notice that if $2n + \alpha + \beta > \sqrt{2n} + 1$ which is equivalent to $\alpha + \beta > \sqrt{2n}(1 - \sqrt{2n}) + 1$, the estimator with the smallest variance is $\hat{\theta}^{SEL}$, but $\sqrt{2n} > 1$ for $n \geq 1$, and the hyperparameters are positive, hence, $\hat{\theta}^{SEL}$ always has the smallest variance and for moderate or large sample sizes, the differences between $Var_\theta[\hat{\theta}^{SEL}]$ and $Var_\theta[\hat{\theta}^{QEL}]$ are negligible. Therefore, from the frequentist point of view, the proposed estimators are more precise than the conventional MLE and the differences tend to be more relevant for small sample sizes. Figure 3 shows the behavior of frequentist variances across the sample space for all the estimators in the bialleic case. In that example $n = 691, \alpha = 240, \beta = 240$.

The results are very similar for the multiallelic case. Estimator variances are:

$$Var\left[\left(\widehat{\boldsymbol{\theta}}_i^{ML}\right)_j\right] = \frac{\theta_{j_i}(1-\theta_{j_i})}{2n}$$

$$Var\left[\left(\widehat{\boldsymbol{\theta}}_i^{M-SEL}\right)_j\right] = \frac{2n\theta_{j_i}(1-\theta_{j_i})}{(2n+\alpha^*)^2}$$

$$Var\left[\left(\widehat{\boldsymbol{\theta}}_i^{M-Minimax_1}\right)_j\right] = \frac{\theta_{j_i}(1-\theta_{j_i})}{\left(\sqrt{2n}+1\right)^2}$$

$$Var\left[\left(\widehat{\boldsymbol{\theta}}_i^{M-Minimax_2}\right)_j\right] = \frac{2n\theta_{j_i}(1-\theta_{j_i})}{(2n+n_i-1)^2}$$

If $\alpha_{j_i} + y_{j_i} - 1 > 0$, then:

$$Var\left[\left(\widehat{\boldsymbol{\theta}}_i^{M-QEL}\right)_j\right] = \frac{2n\theta_{j_i}(1-\theta_{j_i})}{(2n+\alpha^*-1)^2}$$

Since $n_i \geq 3$ and $\alpha^* > 0$, $\left(\widehat{\boldsymbol{\theta}}_i^{M-SEL}\right)_j, \left(\widehat{\boldsymbol{\theta}}_i^{M-Minimax_1}\right)_j$ and $\left(\widehat{\boldsymbol{\theta}}_i^{M-Minimax_2}\right)_j$ have uniformly smaller variance than the MLE and if $\alpha^* > 1, \left(\widehat{\boldsymbol{\theta}}_i^{M-QEL}\right)_j$ also has smaller variance than the MLE.

3.2 Numerical example

To illustrate the methodology, a numerical example is presented. Suppose that in a given sample of size $n = 1382$, three biallelic loci are studied. The three possible genotypes at each locus are denoted as $AA_i, AB_i$ and $BB_i, i = 1,2,3$. The target is to obtain point estimators of the frequencies of the $B_i$ alleles $\boldsymbol{\theta} = (\theta_1, \theta_2, \theta_3)'$. The following counts are observed for genotypes $AA_i, AB_i$ and $BB_i$ respectively: 0, 0, 1382 for locus 1; 1245, 132, 5 for locus 2; and 189, 644, 549 for locus 3. As in any Bayesian analysis, prior knowledge can help to set the values of hyperparameters. On the other hand, in the absence of such knowledge, objective priors can be used or an empirical Bayes approach can be implemented to estimate these unknown quantities. To illustrate how hyperparameters could be defined, suppose that the population under study is composed of subgroups. Each subgroup exchanges individuals with the population at a constant rate $m$ and linear pressure is assumed (Kimura and Crow, 1970). The interest is to estimate $\boldsymbol{\theta}$ in a given subgroup. Under this scenario, allelic frequencies at a given locus follow a beta distribution with parameters: $\alpha = 4N_e m p_I, \beta = 4N_e m(1 - p_I)$, where $N_e$ is the effective size of the subgroup and $p_I$ is frequency of the reference allele among the immigrants. Assume that based on knowledge of the population (e.g., preliminary data), it is believed that $N_e = 150$, $m = 0.8$ and that following Kimura and Crow (1970, page 438) it is assumed that the immigrants are a random sample of the complete population, which implies that $p_I$ can be assumed to be constant and equal to the prior population mean. Suppose that information about $p_I$ is available only for two of the loci and it is equal to 0.8 and 0.5 respectively. For locus 3 there is no previous information and therefore a uniform (0,1) prior is used. Using this information, the following estimators are obtained:

$$\widehat{\boldsymbol{\theta}}^{SEL} = \left(\hat{\theta}_1^{SEL}, \hat{\theta}_2^{SEL}, \hat{\theta}_3^{SEL}\right)' = (0.9260, 0.0825, 0.6302)'$$

$$\widehat{\boldsymbol{\theta}}^{QEL} = \left(\hat{\theta}_1^{QEL}, \hat{\theta}_2^{QEL}, \hat{\theta}_3^{QEL}\right)' = (0.9263, 0.0822, 0.6302)'$$

$$\widehat{\boldsymbol{\theta}}^{Minimax_1} = \left(\hat{\theta}_1^{Minimax_1}, \hat{\theta}_2^{Minimax_1}, \hat{\theta}_3^{Minimax_1}\right)' = (0.9907, 0.0597, 0.6278)'$$

$$\widehat{\boldsymbol{\theta}}^{Minimax_2} = \left(\hat{\theta}_1^{Minimax_2}, \hat{\theta}_2^{Minimax_2}, \hat{\theta}_3^{Minimax_2}\right)' = ("DNE", 0.0514, 0.6302)'$$

where "DNE" stands for "does not exist". For the first locus, genotypes $AA_1$ and $AB_1$ are not observed, this is why $\hat{\theta}_1^{Minimax_2}$ does not exist. Moreover, allele $A_1$ was not observed, but the estimators $\hat{\theta}_1^{SEL}, \hat{\theta}_1^{QEL}$ and $\hat{\theta}_1^{Minimax_1}$ are not equal to one (the MLE) because they contain prior information. This is relevant because of the fact that if an allele is not observed in a sample, this does not imply that it does not exist in the population. In addition, when working with SNP chips or other sort of molecular markers, genotyping errors could cause rare allelic variants not to be identified. It has to be taken into account that this situation happens only when some allele is not observed and the appropriate hyperparameter ($\alpha$ for allele A and $\beta$ for allele B) is greater than one. Under different biological scenarios, such as those discussed in Wright (1930; 1937) and Kimura and Crow (1970), the hyperparameters $\alpha$ and $\beta$ will be greater than one for populations with moderate or large effective size. Notice that the largest differences among estimators where for locus 2, where there were low counts of the reference allele. In addition, given the migration rate and allelic frequencies in the immigrants, the hyperparameters are linear functions of the effective population size. Thus, because of the results discussed in section 3.1, under the model assumed in this example, the larger the effective population size, the larger the differences between $\widehat{\boldsymbol{\theta}}^{SEL}, \widehat{\boldsymbol{\theta}}^{QEL}$ and the MLE. Also, the larger the $N_e$, the larger the reduction in variance of these two estimators relative to the variance of the MLE.

## 4. Discussion

The most widely used point estimator of allele frequencies is the MLE, which can be derived using a multinomial distribution for counts of individuals in each genotype or equivalently the counts of alleles and it corresponds to the sample mean. For biallelic loci, the minimaxity property of the MLE was, at least to our knowledge, an unknown fact in the area of quantitative genetics. In addition, it was also shown that this is a Bayes estimator under SEL and a uniform(0,1) prior. It is important to notice that the minimaxity of the estimator holds only when both alleles are observed, that is, $x_{2_i} + 2x_{3_i} > 0$, $2x_{1_i} + x_{2_i} > 0 \ \forall \ i = 1,2,\dots,k$. This situation is not rare when working with actual genotypic data sets; for example, data from single nucleotide polymorphism chips. Under this condition, the estimator is also an unbiased Bayes estimator. For single-parameter estimation problems, Bayesness and unbiasedness are properties combined in a theorem due to Blackwell and Girshick (1954) which establishes that for parametric spaces corresponding to some open interval of the reals, under QEL, and finite expectation of $w(\theta)$, the Bayesian risk of an unbiased Bayes estimator is zero, which is an appealing property. Here, the theorem does not hold because by basic properties of the Beta distribution (Casella and Berger, 2002) for $\alpha = 1, \beta = 1$, and the particular choice of $w(\theta)$ that was used here, $E[w(\theta)]$ is not finite. Among all the derived estimators, $\hat{\theta}^{Minimax_2}$ and its multivariate version $\widehat{\boldsymbol{\theta}}^{Minimax_2}$ were the only unbiased estimators. Let $B_\theta(\cdot)$ denote the bias of a given estimator. The following are the biases of the estimators derived here:

$$B_\theta(\hat{\theta}^{SEL}) = \frac{-\theta(\alpha + \beta) + \alpha}{2n + \alpha + \beta}$$

$$B_\theta(\hat{\theta}^{QEL}) = \frac{-\theta(\alpha + \beta - 2) + \alpha - 1}{2n + \alpha + \beta - 2}$$

$$B_\theta(\hat{\theta}^{Minimax_1}) = \frac{1 - 2\theta}{2(\sqrt{2n} + 1)}$$

$$B_\theta\left(\hat{\theta}^{Minimax_2}\right) = 0$$

$$B_{\theta_{j_i}}\left(\left(\hat{\boldsymbol{\theta}}_i^{M-SEL}\right)_j\right) = \frac{\alpha_{j_i} - \alpha^* \theta_{j_i}}{2n + \alpha^*}$$

$$B_{\theta_{j_i}}\left(\left(\hat{\boldsymbol{\theta}}_i^{M-QEL}\right)_j\right) = \frac{\alpha_{j_i} - 1 - \theta_{j_i}(\alpha^* - 1)}{2n + \alpha^* - 1}$$

$$B_{\theta_{j_i}}\left(\left(\hat{\boldsymbol{\theta}}_i^{M-Minimax_1}\right)_j\right) = \frac{\frac{1}{n_i} - \theta_{j_i}}{\sqrt{2n} + 1}$$

$$B_{\theta_{j_i}}\left(\left(\hat{\boldsymbol{\theta}}_i^{M-Minimax_2}\right)_j\right) = \frac{-\theta_{j_i}(n_i - 1)}{2n + n_i - 1}$$

The Bayes decision rules derived under QEL depend on $x_{2_i} + 2x_{3_i} + \alpha - 1$ and $2x_{1_i} + x_{2_i} + \beta - 1$. At locus $i$, when the "reference" allele is fixed and $\beta_i \leq 1$, that is, $2x_{1_i} + x_{2_i} + \beta_i - 1 \leq 0$, $R\left(\theta, \hat{\theta}^{QEL}\right) = \frac{1-\theta_i}{\theta_i}$ which is zero when $\theta_i$ is one and tends to infinite as $\theta_i$ approaches zero. Similarly, when the "reference" allele is not observed and $\alpha_i \leq 1$, $R\left(\theta, \hat{\theta}^{QEL}\right) = \frac{\theta_i}{1-\theta_i}$, which is zero when $\theta_i$ is zero and tends to infinite and $\theta_i$ approaches one. Using these results, the k loci situation can be easily analyzed since the loss is additive and hence the risk too. If a set of loci have fixed alleles, the contributions to the risk function in the remaining alleles is finite, and if some of the loci with fixed alleles meet the conditions under which their contributions to the risk tend to infinite, then the risk will tend to infinite. Notice that this can be easily avoided by choosing hyperparameters with values greater than one.

It was found that the risk function under KLL does not have a closed form since it involves finite summations without closed forms. However, this does not prevent the computation of that risk function. Markov chain Monte Carlo methods could be used to compute $E_\theta[\ln(Y_1 + \alpha)]$ and $E_\theta[\ln(Y_2 + \beta)]$ and hence, the risk function could be computed.

In the multiallelic scenario, similar to the biallelic case, when the loss is QEL, the existence of a minimax estimator depends on the condition $y_{j_i} > 0, \forall j_i = 1,2,\ldots,n_i, \forall i = 1,2,\ldots,k$. This means that all allelic variants have to be observed in order to have a minimax estimator under the particular QEL used here. When this condition does not hold for all loci, that is, at least one of them (e.g., $i$) is such that the $j_i^{th}$ allele is not observed, and the corresponding hyperparameter is smaller or equal than one, then the estimator is zero and the risk contribution of this allele is $\theta_{j_i}$. Therefore, in this case the risk does not tend to infinite as was the case for the biallelic scenario; this is due to the fact that the loss function was not the same.

It has to be considered that QEL is a flexible loss function in the sense that the only requirement for $w(\theta)$ is to be positive. Thus, several Bayes estimators can be found by varying this function and possibly, applying theorem 1, other Minimax estimators could be found. The forms of $w(\theta)$ used here for the biallelic and multiallelic case were chosen to cancel with similar expressions depending on $\theta$ during the derivation of the risk functions.

For all decision rules derived from SEL, the form of the risk functions shows that they converge to zero as $n \to \infty$. For QEL, it depends on the possible fixation or absence of a given allele at some loci and the value of the hyperparameters. When all hyperparameters are greater than one, all the derived risk functions converge to zero as $n \to \infty$. When some alleles are fixed (biallelic case) or some are not observed (general case) and the hyperparameters corresponding to their frequencies are smaller or equal to one, the result does not hold.

Admissibility holds for all the estimators derived from SEL while for QEL, if the hyperparameters are greater than one or all allelic variants at each locus are observed (which implies no fixed alleles) the Bayes estimators derived from this loss are also admissible.

Moreover, if all alleles are observed it is possible to obtain admissible minimax estimators from QEL.

Regarding the behavior of the proposed decision rules, the general expressions for the ratios of estimators derived here may be used to have an insight of settings under which the estimators could show large differences and when they do not. For example, estimators derived under SEL differ from the MLE for low counts of the reference alleles and large values of the hyperparameters. From the frequentist point of view, the estimators proposed here always have a uniformly smaller variance than the MLE, except for those derived from QEL which require conditions over the sum of the hyperparameters to meet this property: $\alpha + \beta > 2$ in the biallelic case and $\alpha^* > 1$ in the multiallelic case. However, in many practical applications (as the one provided in the example) these conditions would be satisfied. Although there exists an algebraic reduction of variance, in some situations it could be negligible. For estimators derived under SEL and QEL, the reduction in variance increases as the hyperparameters increase. Also, the reduction in frequentist variances are more marked for small sample sizes. For large sample sizes differences between estimators can still be considerable (see Figure 1).

The impact of using these estimators on each of their applications can be assessed either empirically or theoretically and this is an area for further research. An application in genome-wide prediction or genomic selection (Meuwissen et al., 2001), a currently highly studied area, could be of interest because when both genotypes and their effects are treated as independent random variables, the variance of the distribution of a breeding value is affected by differences in allelic frequencies, by the variance of the distribution of marker effects, and by the level of heterozygosity which is computed using allelic frequencies (Gianola et al., 2009). Other relevant fields where the performance of alternative point estimators of allelic frequencies could be evaluated are the computation of marker-based additive relationship matrices (VanRaden, 2008) and the detection of selection signature using genetic markers (Gianola et al., 2010).

## 5. Conclusion

From the statistical point of view, estimators combining desired statistical properties as Bayesness, minimaxity and admissibility were found and it was shown that for biallelic loci, in addition to the unbiasedness property of the usual estimator, it is also minimax and admissible (provided that all alleles are observed).

Beyond their statistical properties, the estimators derived here have the appealing property of taking into account random variation in allelic frequencies, which is more congruent with the reality of finite populations exposed to evolutionary forces.

**Acknowledgements**

C. A. Martínez acknowledges Fulbright Colombia and "Departamento Adiministrativo de Ciencia, Tecnología e Innovación" COLCIENCIAS for supporting his PhD and Master programs at the University of Florida through a scholarship. Authors acknowledge professor Malay Ghosh from the Department of Statistics of the University of Florida for useful comments.

## Figure captions

**Figure 1** Behavior of ratios: $\delta^{SEL}$ (SEL), $\delta^{QEL}$ (QEL) and $\frac{\hat{\theta}^{SEL}}{\hat{\theta}^{Minimax_1}}$ (S:M1) as functions of $\beta$ for sample sizes 1382 (case A) and 691 (case B) and scenarios 1 and 2 (In scenario 2 SEL and QEL are almost overlapped)

**Figure 2** Behavior of the ratio $\delta^{Minimax_1}$ as a function of the observed frequency of the reference allele B for four different sample sizes (n).

**Figure 3** Frequentist variances of the proposed estimators and the MLE for the biallelic case (Variances of SEL and QEL are almost overlapped)

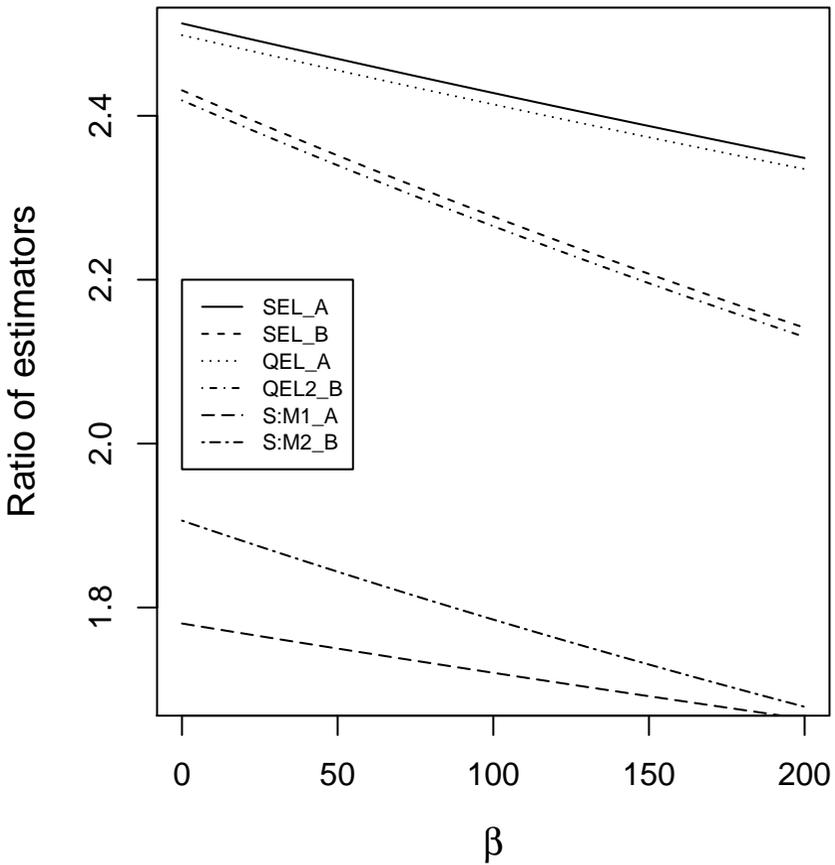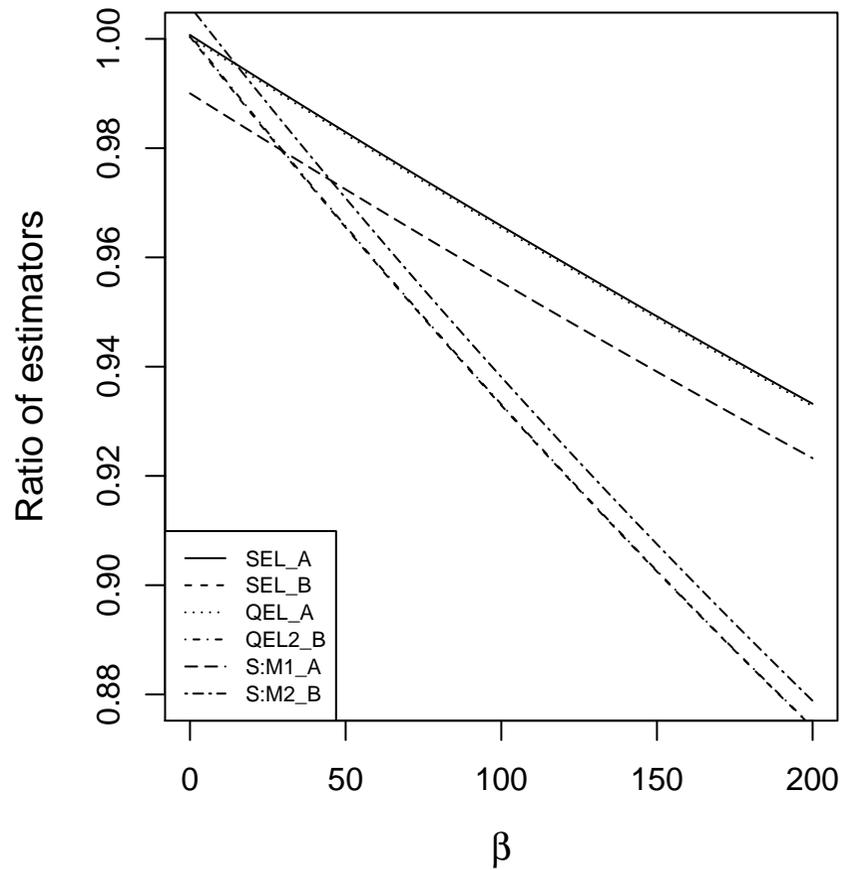

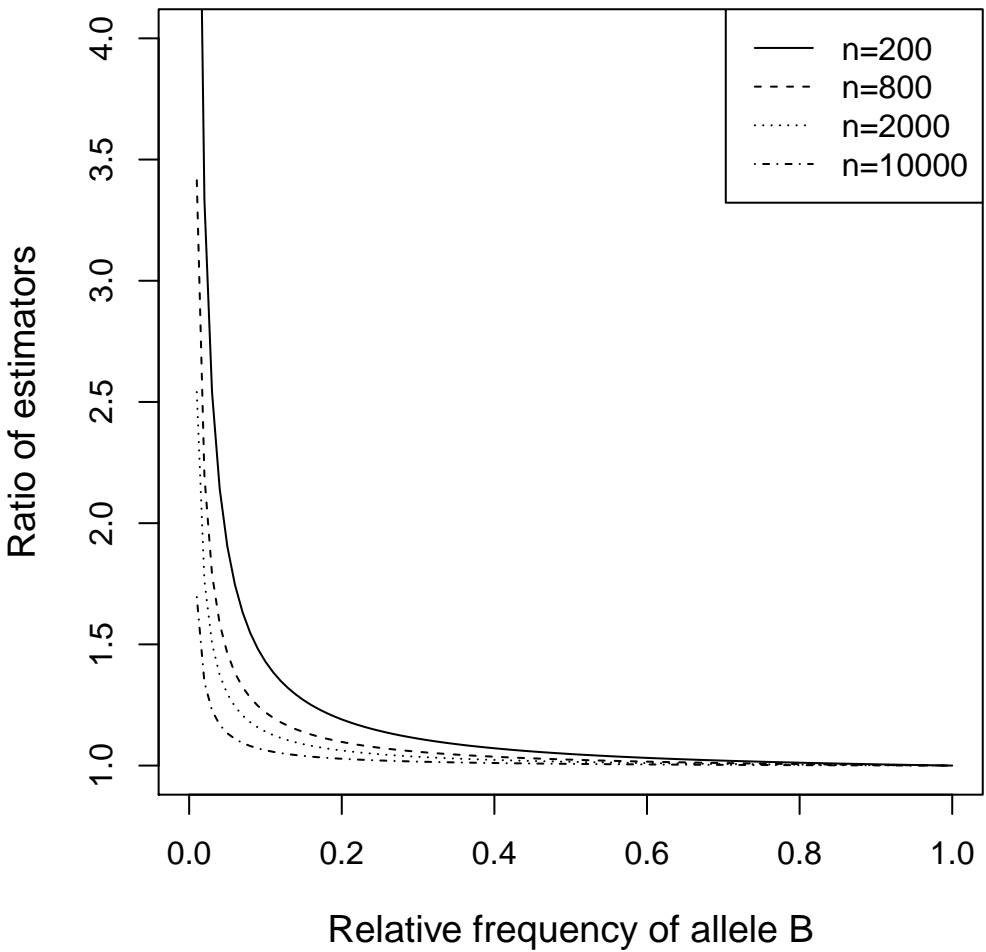

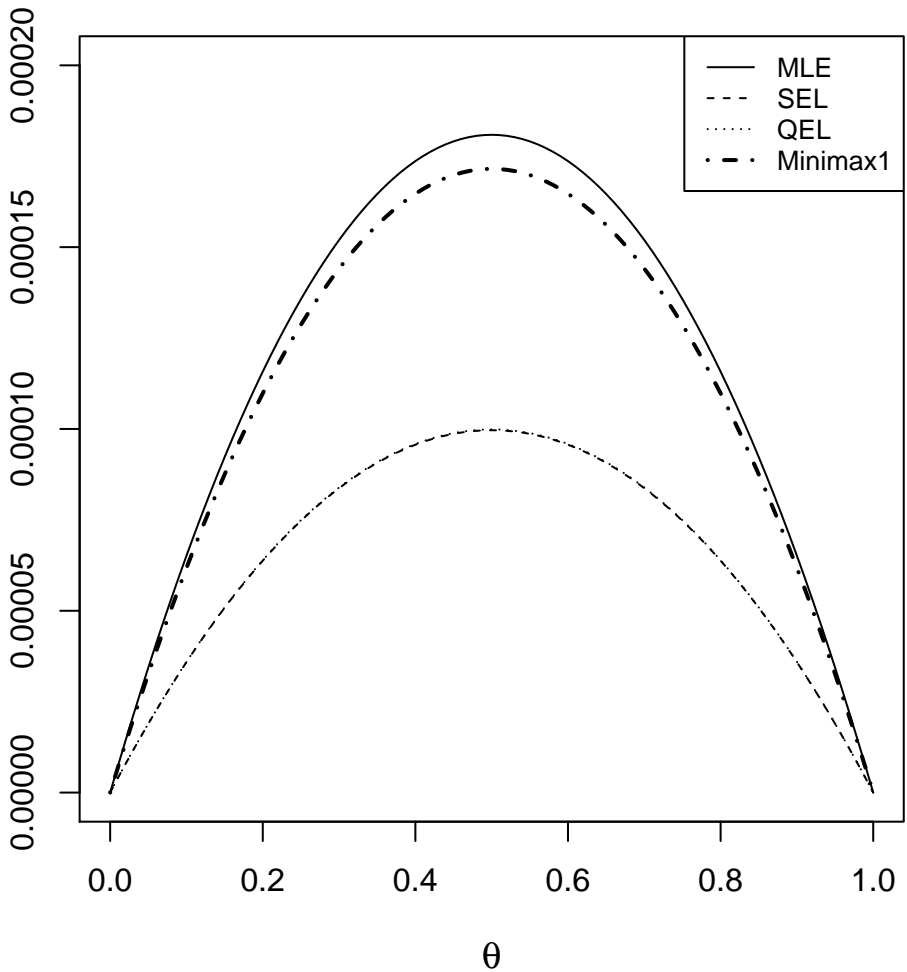